\documentclass[epj]{svjour}
\usepackage{graphics}
\newcommand{\rgn}{($\gamma$,n)}

\newcommand{\rga}{($\gamma$,$\alpha$)}
\newcommand{\rag}{($\alpha$,$\gamma$)}

\newcommand{\oag}{$^{16}$O\rag $^{20}$Ne}
\newcommand{\nega}{$^{20}$Ne\rga $^{16}$O}

\begin{document}
\title{
Relation between the $^{16}$O($\alpha$,$\gamma$)$^{20}$Ne reaction 
  and its reverse $^{20}$Ne($\gamma$,$\alpha$)$^{16}$O reaction in
  stars and in the laboratory
}
\author{
P.~Mohr\inst{1} \thanks{\emph{email:} WidmaierMohr@compuserve.de} \and 
C.~Angulo \inst{2} \and
P.~Descouvemont \inst{3} \and
H.~Utsunomiya \inst{4}
}                     
%
%
\institute{
Diakoniekrankenhaus, D-74523 Schw\"abisch Hall, Germany \and
Centre de Recherches du Cyclotron, UCL, Louvain-la-Neuve, Belgium \and
PNTPM, Universit\'e Libre de Bruxelles, Brussels, Belgium \and
Konan University, Kobe, Japan
}
\date{Received: date / Revised version: date}
%
\abstract{ The astrophysical reaction rates of the \oag\ capture
reaction and its inverse \nega\ photodisintegration reaction are given
by the sum of several narrow resonances and a small direct capture
contribution at low temperatures. Although the thermal population of
low lying excited states in $^{16}$O and $^{20}$Ne is extremely small,
the first excited state in $^{20}$Ne plays a non-negligible role for
the photodisintegration rate. Consequences for experiments with
so-called quasi-thermal photon energy distributions are discussed.
} 
\titlerunning{
Relation between $^{16}$O($\alpha$,$\gamma$)$^{20}$Ne and its reverse
$^{20}$Ne($\gamma$,$\alpha$)$^{16}$O in stars and in the laboratory
}
\maketitle
\section{Introduction}
\label{intro}
\begin{figure}
\resizebox{0.5\textwidth}{!}{
  \includegraphics{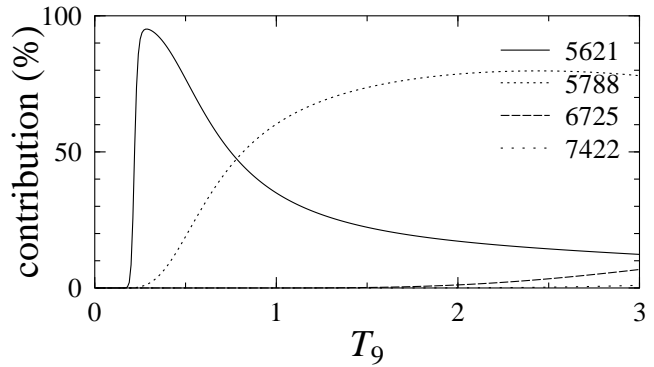}
}
\caption{Contribution of individual resonances to \oag\ reaction
  rate. The resonances are labelled by their energies $E_R$ in keV
  (see also Table \ref{tab:level}). At very low temperatures below
  $T_9 = 0.2$ direct capture is dominating.  
}
\label{fig:contrib_for}
\end{figure}

The small reaction rate of the \oag\ capture reaction blocks the
reaction chain 3\,$\alpha$ $\rightarrow$ $^{12}$C\rag \oag\ in helium
burning at typical temperatures around $T_9 = 0.2$ ($T_9$ is the
temperature in billion degrees K). Its inverse \nega\
photodisintegration reaction is one of the key reactions in neon
burning at higher temperatures around $T_9 = 1 - 2$ \cite{Thi85}. Only
at very low temperatures below $T_9 = 0.2$ the \oag\ reaction rate is
dominated by direct capture. At higher temperatures the reaction rate
is given by the sum of several resonances \cite{NACRE}. The
contributions of several low-lying resonances to the reaction rate of
the \oag\ reaction are shown in Fig.~\ref{fig:contrib_for}.
The properties of three bound states in $^{20}$Ne and four selected
low-lying resonances are listed in Table \ref{tab:level}.

Usually, the reaction rates of inverse photodisintegration reactions
are calculated from the capture rates using the detailed-balance
theorem which is only valid if all nuclei involved are fully thermalized
(see e.g.\ \cite{NACRE}). In the case of light nuclei where the first
excited states are located at relatively high energies one finds very
small occupation probabilities for these excited states. The scope of
this paper is to analyze the relation between the \oag\ and
\nega\ reaction rates for the particular case of high-lying first
excited states ($^{16}$O: $E_x = 6049$\,keV, $0^+$; $^{20}$Ne: $E_x =
1634$\,keV, $2^+$). Furthermore, we discuss potential
experiments using so-called quasi-thermal photon spectra
\cite{Mohr00,Uts05} where the target nucleus is always in its ground
state.

\begin{table}
\setlength{\tabcolsep}{0.9ex}
\caption{Properties of levels in $^{20}$Ne below and above the
  $^{16}$O-$\alpha$ threshold at $Q = 4730$\,keV \cite{Audi03} (from
  \cite{NACRE,Til98}).}
\label{tab:level}       
\begin{tabular}{rcrcccc}
\hline\noalign{\smallskip}
$E_x$ & $J^\pi$ & $E_R^\alpha$ & $(\omega \gamma)$
\cite{Til98} & $(\omega \gamma)$ \cite{NACRE} & $B^0$ & $B^{1634}$
\\ (keV) & & (keV) & (meV) & (meV) & (\%) & (\%) \\
\noalign{\smallskip}\hline\noalign{\smallskip}
0    & $0^+$ & $-$       & $-$      & $-$      & $-$ & $-$\\
1634 & $2^+$ & $-$       & $-$      & $-$      & $100$ & $-$ \\
4248 & $4^+$ & $-$       & $-$      & $-$      & $0$ & $100$ \\
5621 & $3^-$ & $892$     & $1.7(3)$ & $1.9(3)$ & $7.6(10)$ & $87.6(10)$ \\
5788 & $1^-$ & $1058$    & $17(3)$  & $23(3)$  & $18(5)$ & $82(5)$ \\
6725 & $0^+$ & $1995$    & $71(12)$ & $74(9)$  & $0$ & $100$ \\
7422 & $2^+$ & $2692$    & $146(19)$& $160(20)$& $\le 9.4(14)$ &
$\ge 90.6(14)$ \\ 
\noalign{\smallskip}\hline
\end{tabular}
\end{table}

\section{Resonant reaction rates}
\label{sec:rates}
Because of the dominance of resonances in the rates of the \oag\ and
\nega\ reactions we restrict ourselves to the discussion of
resonances. The reaction rate
$\left\langle \sigma \, v \right\rangle $ of a
Breit-Wigner resonance with 
\begin{equation}
\sigma_{BW}(E) \, = \, \frac{\pi}{k_\alpha^2} \, \omega \,
  \frac{\Gamma_\alpha \Gamma_\gamma}{(E-E_R^\alpha)^2+\Gamma^2/4}
\label{eq:bw_for}
\end{equation}
is given by:
\begin{equation}
\left\langle \sigma \, v \right\rangle 
\, = \, \hbar^2 \left( \frac{2 \pi}{\mu kT}
\right)^{3/2} \, (\omega \gamma) \,
\exp{\left(\frac{-E_R^\alpha}{kT}\right)}
\label{eq:rate_for}
\end{equation}
with the reduced mass $\mu$, the resonance energy $E_R^\alpha$ in the
c.m.-system, and the resonance strength 
\begin{equation}
(\omega \gamma) = \omega \,
\frac{\Gamma_\alpha \Gamma_\gamma}{\Gamma}
\label{eq:strength}
\end{equation}
and the statistical factor
\begin{equation}
\omega =
\frac{(2J_R+1)}{(2J_1+1) (2J_2+1)}
\label{eq:omega}
\end{equation}
$\Gamma_\alpha$ is the $\alpha$ decay width of the resonance,
\begin{equation}
\Gamma_\gamma = \sum_b \Gamma_\gamma^b
\label{eq:gamma_gamma}
\end{equation}
the total radiation width (summed over all partial radiation widths
$\Gamma_\gamma^b$ to bound states $b$), $\Gamma
= \Gamma_\alpha + \Gamma_\gamma$ the total decay width, and 
\begin{equation}
B^b = \Gamma_\gamma^b/\Gamma_\gamma
\label{eq:branch}
\end{equation}
is the $\gamma$-ray branching ratio to a bound state $b$.

The total astrophysical reaction rate is obtained by adding the
contributions of all resonances and of a small direct contribution. In
the case of the \oag\ reaction, all contributions in the
laboratory (with the target in the ground state) and under stellar
conditions are identical up to high temperatures because of the high
lying first excited state in $^{16}$O \cite{NACRE}.

The cross section of the \nega\ photodisintegration reaction is
directly related to the capture cross section by time-reversal
symmetry:
\begin{equation}
\frac{\sigma_{3\gamma}(E_\gamma)}{\sigma_{12}(E_{\alpha})} = 
	\frac{\lambda^2_\gamma}{\lambda^2_{12}} \,
	\frac{(2J_1+1) (2J_2+1)}{2\,(2J_3+1)}
\label{eq:reverse}
\end{equation}
The cross section of a Breit-Wigner resonance for the capture reaction
to a bound state $b$ is given by 
\begin{equation}
\sigma_{12}(E_{\alpha}) = B^b \, \times
\, \sigma_{BW}(E)
\label{eq:BW_br}
\end{equation}
with the branching ratio $B^b$ as defined in
Eq.~(\ref{eq:branch}). Using Eq.~(\ref{eq:reverse}), the thermal
photon density from the Planck law
\begin{equation}
n_\gamma \, (E_\gamma,T) \, dE_\gamma = 
	\frac{1}{\pi^2} \, \frac{1}{(\hbar c)^3} \, 
	\frac{E_\gamma^2}{\exp{(E_\gamma/kT)} - 1} \, dE_\gamma
\label{eq:planck}
\end{equation}
and the definition of the reaction rate
\begin{equation}
\lambda = c \, \int n_\gamma(E_\gamma,T) \, \sigma(E_\gamma) \,
dE_\gamma 
\label{eq:rate_rev}
\end{equation}
one obtains a reaction rate for the target in a defined bound state
$b$ with spin $J_3$:
\begin{equation}
\lambda^b = \frac{1}{\hbar} \, \frac{(2J_1+1) (2J_2+1)}{(2J_3+1)} \,
B^b \, (\omega \gamma) \, \exp{\left(\frac{-E_R^\gamma}{kT}\right)}
\label{eq:rate_rev_ind}
\end{equation}
with the required photon energy 
\begin{equation}
E_R^\gamma = E_R^\alpha + Q - E_x^b
\label{eq:e_phot}
\end{equation}
and $E_x^b$ is the excitation energy of the bound state $b$.
Note that the required photon energy $E_R^\gamma$ for the transition
to a resonance at excitation energy $E_R$ is reduced by the excitation
energy  $E_x^b$ of the bound state $b$ under consideration!

For simplicity we choose the example of the $1^-$ resonance at $E_R =
5788$\,keV in $^{20}$Ne in the following discussion. This resonance
has a strength of $\omega \gamma = 23$\,meV and branching ratios of
$B^0 = 18$\,\% to the ground state and $B^{1634} = 82$\,\% to the
first excited state in $^{20}$Ne at $E_x = 1634$\,keV with $J^\pi =
2^+$ \cite{NACRE,Til98} (see Table \ref{tab:level}). 

For the ground state rate $\lambda^0$ we obtain
\begin{equation}
\lambda^0 = \frac{1}{\hbar} \, 
B^0 \, (\omega \gamma) \, \exp{\left(\frac{-E_R}{kT}\right)}
\label{eq:rate_rev0}
\end{equation}
whereas the rate $\lambda^{1634}$ for $^{20}$Ne in its first excited
state is
\begin{equation}
\lambda^{1634} = \frac{1}{\hbar} \, \frac{1}{5} \,
B^{1634} \, (\omega \gamma) \, \exp{\left(\frac{-(E_R-E_x^{1634})}{kT}\right)}
\label{eq:rate_rev1634}
\end{equation}
Note the factor of $1/5$ because of $J^\pi = 2^+$ for the first
excited state at $E_x = 1634$\,keV. Using the thermal occupation
probability ratio
\begin{equation}
n^{1634}/n^0 =
\frac{(2J^{1634}+1)}{(2J^0+1)}\exp{\left(\frac{-E_x^{1634}}{kT}\right)}
\label{eq:boltzmann}
\end{equation}
we can calculate the total astrophysical
reaction rate $\lambda^\star$ for this single narrow $1^-$ resonance:
\begin{eqnarray}
\lambda^\star & \approx & \frac{1}{\hbar} \, (\omega \gamma) \left[
B^{0} \, \exp{\left(\frac{-E_R}{kT}\right)} \, + \right. \nonumber \\
& & \, \, \, \, \left. \frac{1}{5} \, B^{1634} \,
\exp{\left(\frac{-(E_R-E_x^{1634})}{kT}\right)} \, 5 \,
\exp{\left(\frac{-E_x^{1634}}{kT}\right)} \right] \nonumber \\ 
& = & \frac{1}{\hbar} \, (\omega \gamma) \, \left( B^0 + B^{1634} \right) \, 
\exp{\left(\frac{-E_R}{kT}\right)} \nonumber \\
& = & \frac{1}{\hbar} \,  (\omega \gamma) \,
\exp{\left(\frac{-E_R^\alpha}{kT}\right)} \,
\exp{\left(\frac{-Q}{kT}\right)}
\label{eq:rate_rev_star}
\end{eqnarray}

Comparing Eqs.~(\ref{eq:rate_for}) and (\ref{eq:rate_rev_star}), one
finds the relation between the capture rate $\left\langle
\sigma v \right\rangle$ and the photodisintegration rate
$\lambda^\star$:
\begin{equation}
\frac{\lambda^\star}{\left\langle \sigma v \right\rangle} 
= \left(\frac{\mu \, kT}{2\pi
  \hbar^2}\right)^{3/2} \, \exp{\left(\frac{-Q}{kT}\right)} 
\label{eq:rate_ratio}
\end{equation}
This is identical to the result of the detailed balance theorem for
$J^\pi(\alpha) = J^\pi(^{16}{\rm{O}}) = J^\pi(^{20}{\rm{Ne}}) = 0^+$
and $G(\alpha) = G(^{16}{\rm{O}}) = G(^{20}{\rm{Ne}}) = 1$ where $G$
are the temperature-dependent normalized partition functions as e.g.\
defined in \cite{NACRE}. Following \cite{NACRE}, $G(T)$ do not deviate
more than $1$\,\% from unity for $\alpha$, $^{16}$O, and $^{20}$Ne up
to $T_9 = 3$.

\section{Discussion}
\label{sec:disc}
There are several interesting aspects which arise from the
calculations in Sect.~\ref{sec:rates}. As already pointed out, the
first excited states of $^{16}$O and $^{20}$Ne have relatively high
excitation energies. E.g., at $T_9 = 1$ this leads to occupation
probabilities of $3 \times 10^{-31}$ for the $0^+$ state in $^{16}$O
at $E_x = 6049$\,keV and $3 \times 10^{-8}$ for the $2^+$ state in
$^{20}$Ne at $E_x = 1634$\,keV. The normalized partition functions do
practically not deviate from unity up to $T_9 = 3$ \cite{NACRE}.

The reaction rate $\left\langle \sigma v
\right\rangle$ for the \oag\ capture reaction can be determined from
laboratory experiments because the rates under stellar and laboratory
conditions are identical for this reaction: 
\begin{equation}
\left\langle \sigma v \right\rangle^\star \, = \, \left\langle \sigma
v \right\rangle^{\rm{lab}} 
\label{eq:rate_for_labstar}
\end{equation}

However, this is not the case for the reaction rate of the \nega\
photodisintegration reaction under stellar conditions and in the
laboratory:
\begin{equation}
\lambda^\star \, \ne \, \lambda^{\rm{lab}}
\label{eq:rate_inv_labstar}
\end{equation}
The stellar reaction rate $\lambda^\star$ for a resonance is given by
Eq.~(\ref{eq:rate_rev_star}); the full resonance strength $(\omega
\gamma)$ of the capture reaction contributes to the stellar
photodisintegration rate $\lambda^\star$. The laboratory reaction rate
$\lambda^0$ is reduced by the branching ratio $B^0$ to the ground
state as can be seen from Eq.~(\ref{eq:rate_rev0}). It is interesting
to note that for the chosen example of the $1^-$ resonance at $E_R =
5788$\,keV the laboratory rate $\lambda^0$ is only 18\,\% of the
stellar rate $\lambda^\star$ because of the branching ratio $B^0$. The
first excited state at $E_x = 1634$\,keV in $^{20}$Ne contributes with
82\,\% to the stellar rate $\lambda^\star$ although the thermal
occupation probability is only $3 \times 10^{-8}$ at $T_9 = 1$! The
reason for this surprising contribution can be understood from
Eqs.~(\ref{eq:planck}) and (\ref{eq:rate_rev1634}). The small thermal
occupation probability at $E_x = 1634$\,keV is exactly compensated by
the much higher photon density at the required energy of $E_R^\gamma =
E_R^\alpha + Q - E_x^{1634}$.

\begin{figure}
\resizebox{0.5\textwidth}{!}{
  \includegraphics{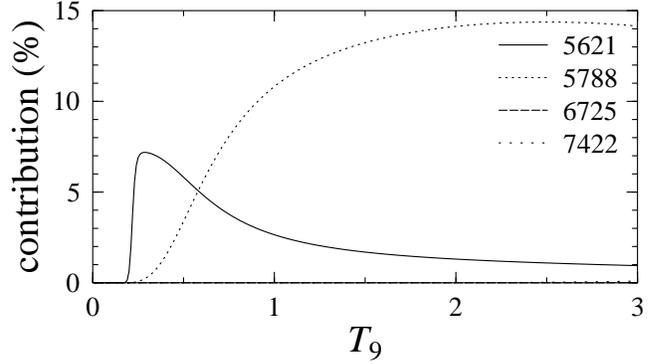}
}
\caption{Ground state contribution $\lambda^{\rm{lab}}/\lambda^\star$
  of individual resonances to the stellar reaction rate
  $\lambda^\star$ of the \nega\ photodisintegration reaction as given
  by detailed balance \cite{NACRE}. The resonances are labelled by
  their energies $E_R$ in keV. There is no contribution from the
  resonance at $E_R = 6725$\,keV because of $B^0 = 0$ for this
  resonance. Note the different scale (compared to
  Fig.~\ref{fig:contrib_for}).  
}
\label{fig:contrib_rev}
\end{figure}

The contribution of the laboratory
reaction rate $\lambda^{\rm{lab}}$ to the stellar rate $\lambda^\star$
(taken from the detailed balance calculation of \cite{NACRE}) is shown
in Fig.~\ref{fig:contrib_rev}. The contributions of these resonances
to the stellar \nega\ reaction rate $\lambda^\star$ are identical to
the contributions in the \oag\ reaction rate which is shown in
Fig.~\ref{fig:contrib_for}.

The ratio $\lambda^\star / \left\langle\sigma v \right\rangle$ does
not depend on individual properties of the respective
resonance. Consequently, Eq.~(\ref{eq:rate_ratio}) is also valid for
the direct capture contribution which becomes relevant at relatively
low temperatures. The detailed balance theorem hence is also
applicable to reactions between nuclei with high-lying first excited
states as e.g.\ the \oag\ and \nega\ reactions.

A further consequence of the relation between capture and
photodisintegration reactions is that the so-called Gamow windows of
both reactions have a strict relation. The Gamow window - the energy
region where a reaction mainly operates - of a capture reaction is
characterized by its position at
\begin{equation}
E_G = 1.22 \left( Z_1^2 Z_2^2 A_{\rm{red}} T_6^2 \right)^{1/3}\,{\rm{keV}}
\label{eq:gamow_e}
\end{equation}
and its width $\Delta$
\begin{equation}
\Delta = 0.749 \left( Z_1^2 Z_2^2 A_{\rm{red}} T_6^5
\right)^{1/6}\,{\rm{keV}}
\label{eq:gamow_delta}
\end{equation}
The corresponding Gamow window of the \rga\ reaction is shifted
by the $Q$-value of the reaction
\begin{equation}
E^\gamma_{\rm{G}} = E^\alpha_{\rm{G}} + Q
\label{eq:gamow}
\end{equation}
The width $\Delta$ remains the same for \rag\ and \rga\ reactions.

In the chosen example of the \oag\ capture and \nega\
photodisintegratiom reactions, one finds e.g.\ at $T_9 = 1$ energies
of $E^\alpha_{\rm{G}} = 1141$\,keV and $E^\gamma_{\rm{G}} = 5871$\,keV
and a width $\Delta = 725$\,keV. As an obvious consequence the $1^-$
resonance at $E_R = 5788$\,keV is dominating around $T_9 = 1$ (see
also Fig.~\ref{fig:contrib_for}). Properties of the Gamow window for
\rga\ reactions have been discussed in further detail in
\cite{Mohr03c,Mohr04c}.

As shown above, the detailed balance theorem holds for the \oag\ and
\nega\ reactions provided that the population of excited states is
thermal according to the Boltzmann statistics. How long does it take
until thermal equilibrium is obtained in such cases where the first
excited states are only weakly polulated? In general, one can read
from the formalism of photoactivation with a constant production rate
(which is approximately fulfilled in the above case), that the time
until equilibrium is reached is a few times the lifetime of the
unstable product. In the case of the $2^+$ state in $^{20}$Ne at $E_x
= 1634$\,keV the mean lifetime is $\tau = 1.05$\,ps
\cite{Til98}. Compared to the timescale of neon burning, which is of
the order of years, thermal equilibrium is reached almost
instantaneously.

\section{Experiments with quasi-thermal photon spectra}
\label{sec:exp}
Using a non-monochromatic photon spectrum, the experimental yield $Y$
from a single resonance in the photodisintegration reaction \nega\ is
given by:
\begin{equation}
Y = N_T n_\gamma(E_R) \left(\frac{\hbar c}{E_R}\right)^2 \pi^2 B^0
\frac{(2J_1+1) (2J_2+1)}{(2J_3+1)} (\omega \gamma)
\label{eq:yield}
\end{equation}
Here $N_T$ is the number of target nuclei, and $n_\gamma(E_R)$ is the
number of incoming photons per energy interval and area at the energy
of the resonance $E_R$. Using realistic numbers for present-day
facilities and thin targets for direct detection of the $\alpha$
particles, one obtains a relatively small yield. E.g., $N_T \approx
10^{17}$, $n_\gamma \approx 10^5/$keV\,cm$^2$\,s, and the weak $1^-$
resonance at $E_R = 5788$\,keV with the properties given in Table
\ref{tab:level}, leads to $Y \approx 0.5/$day. For stronger resonances
the yields are higher by up to two orders of magnitude which makes
experiments difficult but feasible with present-day facilities.

The astrophysical reaction rate $\lambda^\star$ for the \nega\
photodisintegration reaction is well-defined from experimental data of
the \oag\ capture reaction. Nevertheless, experimental yields for
several resonances can be obtained simultaneously in one irradiation
using the quasi-thermal spectrum which will become available at
SPring-8 \cite{Uts05}. Consequently the ratios of all observed
resonances will be measured in one irradiation - provided that the
ground state branching ratios are well-known. This might help to
resolve minor discrepancies between the adopted resonance strengths in
\cite{Til98} and \cite{NACRE}.

There are further interesting experimental properties of the \nega\
reaction which have to be discussed in the following. The analysis of
experiments with non-monochromatic photons (see e.g.\
Refs.~\cite{Mohr00,Vogt01,Vogt02,Sonn04}) requires the precise
knowledge of the absolute number of incoming photons and their energy
dependence. Using a non-monochromatic photon spectrum in combination
with the \nega\ reaction (or any other photodisintegration reaction
which is dominated by narrow resonances) one finds emitted $\alpha$
particles with discrete energies. Because the extremely high-lying
first excited state in $^{16}$O is practically not populated in the
\nega\ reaction, the $\alpha$ energy is given by the difference
between the resonance energy $E_R$ and the $Q$ value of the
reaction. The experimental yield in each of the discrete lines is
directly proportional to the resonance strength $\omega \gamma$, the
ground state branching $B^0$, and the number of incoming photons
$n_\gamma(E_R)$ at resonance energy, as can be read from
Eq.~(\ref{eq:yield}). Provided that the resonance strengths and the
branching ratios are well-known, a \nega\ experiment may help to
determine the properties of the incoming photon spectrum with good
accuracy over a broad energy range. This is especially relevant for
the measurement of \rga\ reaction rates because the relevant energy
region, the so-called Gamow window, is much broader than in the case
of \rgn\ reactions which have mainly been analyzed in the last
years. The intrinsic exponential decrease of photon intensity with
energy for the photon source suggested in \cite{Uts05} may help to
avoid the problems of reproducing the thermal photon distribution over
a broad energy range which arise in the present technique using a
superposition of bremsstrahlung spectra \cite{Mohr00}.

\section{Conclusions}
\label{sec:conc}
The relation between the \oag\ capture reaction and the \nega\
photodisintegration reaction have been discussed in detail. Whereas
the stellar reaction rates of the \oag\ reaction are identical to the
laboratory rates \cite{NACRE}, this is not the case for the
photodisintegration rates under stellar and under laboratory
conditions. Although the thermal population of the first excited state
in $^{20}$Ne remains extremely small at typical temperatures of neon
burning, it nevertheless provides an important contribution to the
reaction rate under stellar conditions. The reason for this surprising
behavior is that the increasing number of thermal photons at the
relevant energy exactly compensates the small thermal occupation
probability according to the Boltzmann statistics. The widely used
detailed balance theorem, which relates reaction rates of capture
reactions to photodisintegration rates, remains valid for the case of
the \oag\ and \nega\ reactions.

Additionally, it has been shown that the \nega\ reaction may be
helpful in calibrating new intense non-monochromatic photon sources as
e.g.\ suggested in \cite{Uts05}. Such a calibration over a broad
energy interval is especially relevant for \rga\ experiments because
of the broader Gamow window.


\end{document}